\documentclass[]{spie}  

\usepackage{amsmath,amsfonts,amssymb}
\usepackage{graphicx}
\usepackage[colorlinks=true, allcolors=blue]{hyperref}
\usepackage{multirow}  
\usepackage{eurosym}
\usepackage[colorlinks=true, allcolors=blue]{hyperref}

\title{Optical turbulence forecast for the European Solar Telescope (EST): the challenge of the day-time regime}

\author[a]{Elena Masciadri}
\author[a]{Alessio Turchi}
\author[a]{Fini Luca}

\affil[a]{INAF-Osservatorio Astrofisico di Arcetri, L.go Enrico Fermi 5, Firenze, Italy}

\authorinfo{Send correspondence to Elena Masciadri - e-mail: elena.masciadri@inaf.it}

\begin{document} 
\maketitle

\begin{abstract}
In this contribution we present preliminary results of a study applied to the Observatories of Roque de Los Muchachos (La Palma) and Teide (Tenerife) in Canary Islands aiming to investigate the possibility to implement an automatic system for the optical turbulence forecasting for the European Solar Telescope (EST) telescope. The study has been carried out in the context of the SOLARNET project and the two mentioned sites were the pre-selected sites for EST. This analysis aimed to investigate the possibility to extend the methodology of the forecast of the optical turbulence developed by our team and performed on top-class ground-based telescopes dedicated to night time observations such as ALTA (@ LBT) and FATE (@ VLT) to the day-time regime. As an ancillary output our very preliminary analysis concludes, that the two sites of Roque de Los Muchachos Observatory (ORM) and Teide Observatory (TO) show comparable characteristics during the day time. Considering that the site of EST has been already identified to be at ORM this can be considered a very useful information from a scientific point of view.
\end{abstract}


\section{INTRODUCTION}
\label{sec:intro}  

As part of the H2020 SOLARNET project, in the last few years we developed a study to investigate the possibility to set-up automatic systems for the
forecast of the optical turbulence in the day-time regimes at the two sites of Roque de los Muchachos
Observatory (ORM, La Palma) and Teide Observatory (TO, Tenerife). At the epoch of the beginning of H2020 SOLARNET  project ORM was the selected site for EST but the definitive decision has been taken on 2021\footnote{\href{https://est-east.eu/component/content/article/14-english/news/1067-est-location-at-roque-de-los-muchachos-observatory-approved?Itemid=659}{https://est-east.eu/component/content/article/14-english/news/1067-est-location-at-roque-de-los-muchachos-observatory-approved?Itemid=659}}. Our study aimed to upgrade to day-time conditions the method of the optical
turbulence forecasting ($C_{N}^{2}$ profiles and integrated astroclimatic parameters) that was developed by our group in INAF 
for night-time mainly in application to the ground-based astronomy. Such a step ahead would allow us to extend the application of our methodology not only to the solar astronomy but also to the field of the free-space optical communication (FCO). We show in this contribution a preliminary analysis towards this final goal that contains interesting and useful informations from a scientific point of view for EST.  

We briefly summarise the context and the tools used for this study. The technique used to forecast the optical turbulence is the numerical one. We use a mesoscale model (Meso-Nh) developed by the Centre National des
Recherches Meteorologiques (CNRM) and Laboratoire d’Aerologie (LA) in Toulouse (France)\cite{lafore1998,lac2018}, and a
dedicated code (Astro-Meso-NH) developed by INAF to forecast the optical turbulence\cite{masciadri1999}. The Astro-Meso-Nh code is in continuous evolution since more than a two decades. The most recent version of the code is described in \cite{masciadri2017}. More recently an hybrid technique have been proposed by our group \cite{masciadri2020} that implies the use of Astro-Meso-Nh plus real-time measurements with an auto-regression technique (that we call AR) that improve forecast at short time scales (a few hours, typically 1h or 2h that are the most relevant one for the flexible scheduling and the science verification). Such an approach has been implemented  in the automatic forecast systems of optical turbulence and atmospheric parameters that have been developed by our group for two among the top-class ground-based telescopes: the Large Binocular Telescope (LBT) in Arizona and the Very Large Telescope (VLT) in Chile. The two tools/projects conceived to provide nightly forecasts above the LBT and the VLT are resepctively ALTA Center\footnote{\href{http://alta.arcetri.inaf.it/}{http://alta.arcetri.inaf.it/}} and FATE\cite{masciadri2023,masciadri2024}. The idea is to move us here towards the day time regime. 

\section{MODEL CONFIGURATION}

\begin{figure*}
\begin{center}
\includegraphics[angle=-90,width=0.8\textwidth]{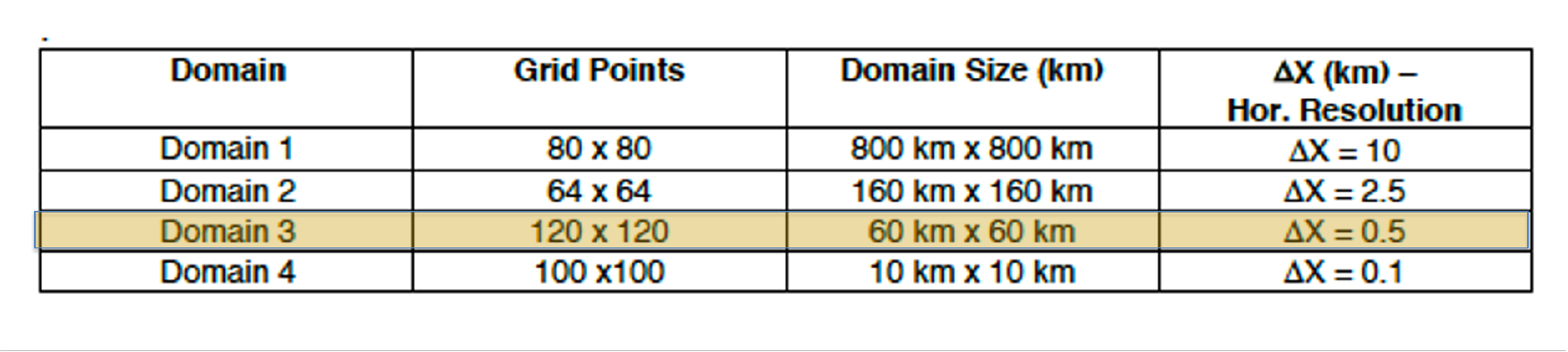}\\
\end{center}
\caption{\label{fig:fig2}  Summary of the geometric characteristics of the imbricated domains used for the grid-nesting configuration. The same configuration has been used for La Palma and Tenerife. The yellow line indicates the domain with the highest horizontal resolution used for the study presented in this contribution. }
\end{figure*}

\begin{figure*}
\begin{center}
\includegraphics[angle=-90,width=\textwidth]{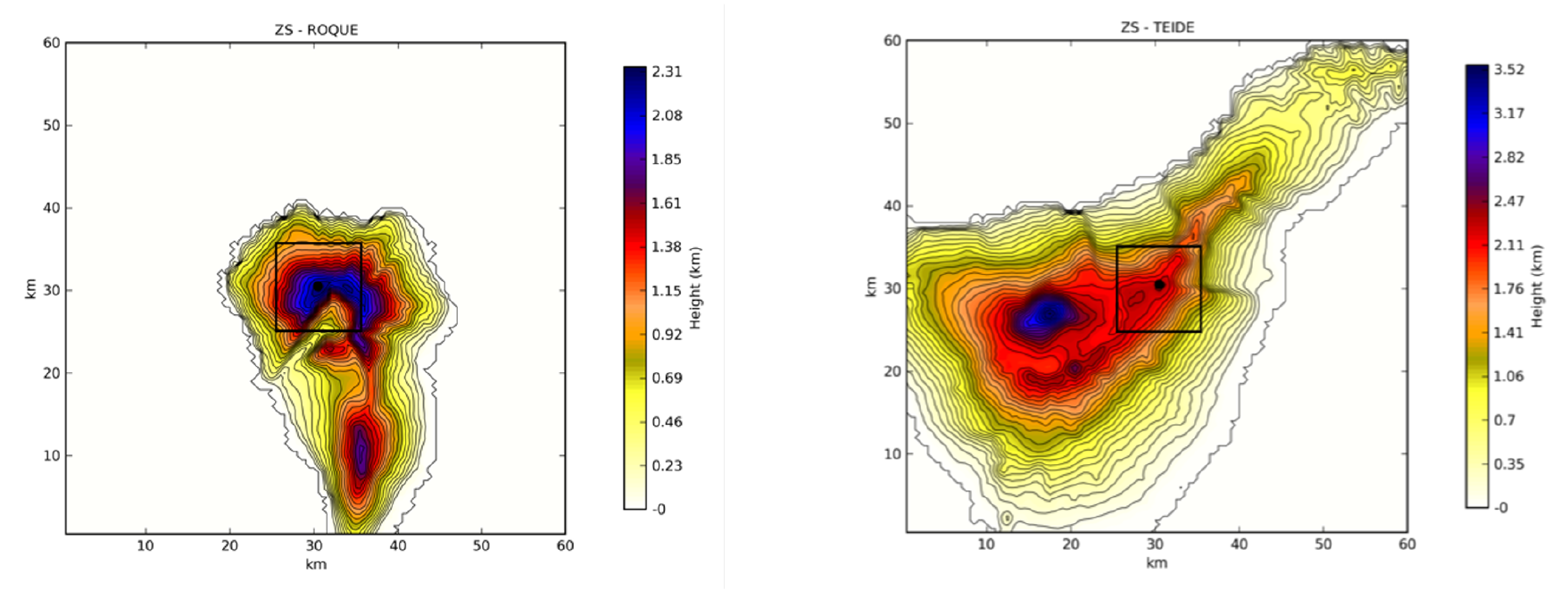}\\
\end{center}
\caption{\label{fig:fig1}  On the left the digital elevation model (orography) of La Palma isle, on the right Tenerife isle. The DOM3 extended on 60~km $\times$ 60~km i.e. 120 $\times$120 grid points with a resolution of 0.5~km. The black squares indicate the DOM4 extended on 10~km $\times$ 10~km i.e. 100 $\times$ 100 grid points. that has not been used in this study.}
\end{figure*}

In Fig.\ref{fig:fig2} are shown the characteristics of the model configuration used for this study. We used the grid-nesting approach\cite{stein2000} centred on the two sites with up to four imbricated domains. The grid-nesting technique permits indeed to use a set of imbricated domains extended on a smaller and smaller surfaces with the innermost domains characterised by the highest resolution.The more external domain with the lowest resolution has been centred above the Gran Telescopio Canarias - GTC (28.756611, -17.89188) at ORM and above the Vacuum Tower Telescope - VTT (28.302389 N, -16.51005 W) at TO. The same set of imbricated domains (same geometric characteristics) has been selected for ORM and TO. The innermost domain has the highest horizontal resolution of 100~m.  The yellow row indicates the highest resolution selected for this study. Fig.\ref{fig:fig1} shows the digital elevation maps of ORM and TO. More precisely it shows the DOM3 extended on 60~km $\times$ 60~km i.e. 120 $\times$ 120 grid points with a resolution of 0.5~km.  The black square centred above the center of the map indicates the extension of the innermost domain with horizontal resolution of 0.1~km. Above ORM is visible the semi-arc of the Caldera characterising the peak of the site hosting sevearal telescopes of class 2-4~m. The digital elevation models (DEM) of ORM is characterised by very sharp slopes in proximity of the summit due to the typical shape of the Caldera. 

The atmospheric model uses 62 vertical levels distributed following the law: the first grid point is 5 m, there is after a logarithm stretching of 20\% up to the height of 3500~m above the ground. Above this threshold the grid size is of around 600~m up to 23.8 km above the ground. The model adjusts by itself the grid size in this part of the atmosphere assuming the number of 62 grid-points in total on the vertical direction.

\subsection{Simulation configuration scheme}

\begin{figure*}
\begin{center}
\includegraphics[width=0.7\textwidth]{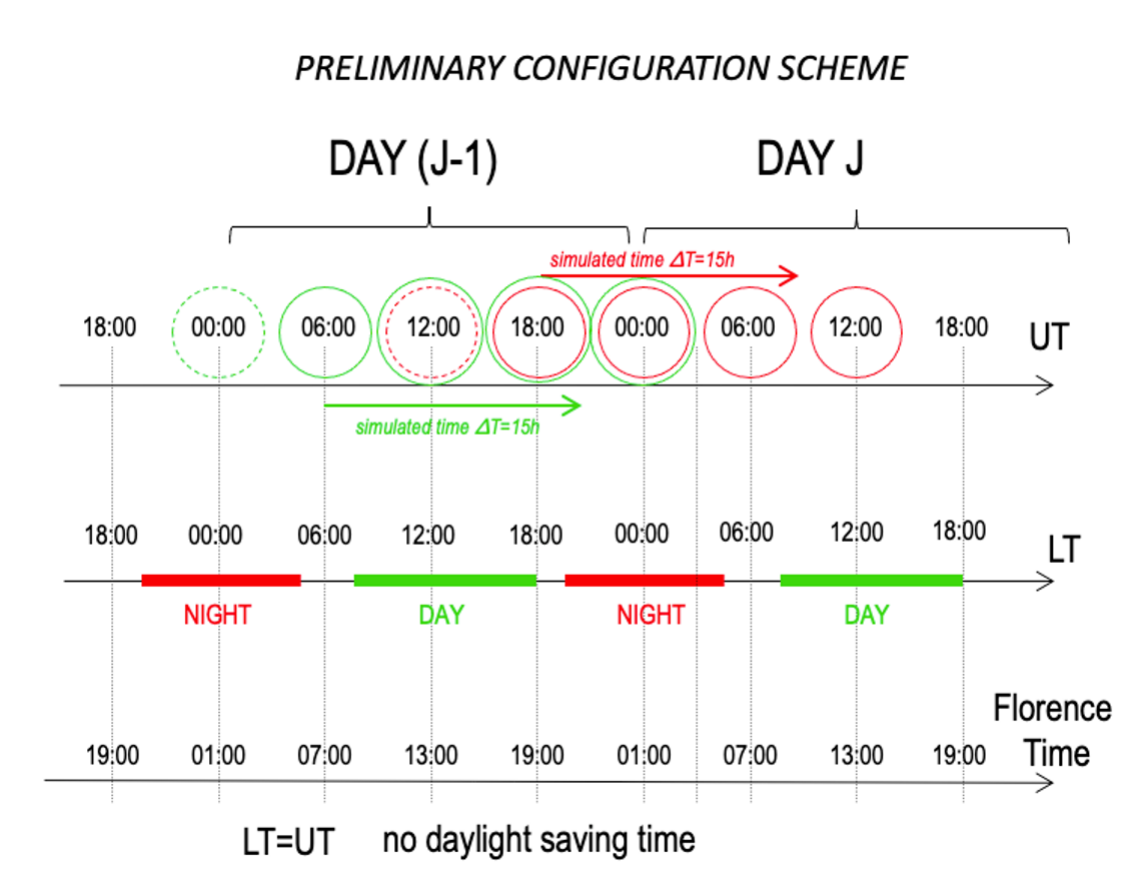}\\
\end{center}
\caption{\label{fig:fig3}  Preliminary operational configuration for the night and day time forecasts. See the text. }
\end{figure*}

The configuration of the model has been conceived to run couples of simulations/forecast (for the day and time period) as indicated in Fig.\ref{fig:fig3}. We conceived a preliminary set-up without the aim to optimise the model performances but to use initialisation data already collected for the purposed of this project. Initialisation and forcing data for the numerical model come from the European Centre for Medium Weather Forecast (ECMWF). For the night time the simulation starts at 18:00 UT of the DAY (J-1) up to 09:00 UT of the DAY J. For the day period the simulation starts at 06:00 UT of the DAY (J-1) and finishes at 21:00 UT of the DAY J. This means to consider a simulated period of 15~hours. Different tests have been performed with 15~hours and 18~hours. Full line circles indicate the initialisation and forcing data. Dashed line circle are alternative solutions for initialisation data that might be tested to refine eventually the configuration.

\begin{figure*}
\begin{center}
\includegraphics[angle=-90,width=\textwidth]{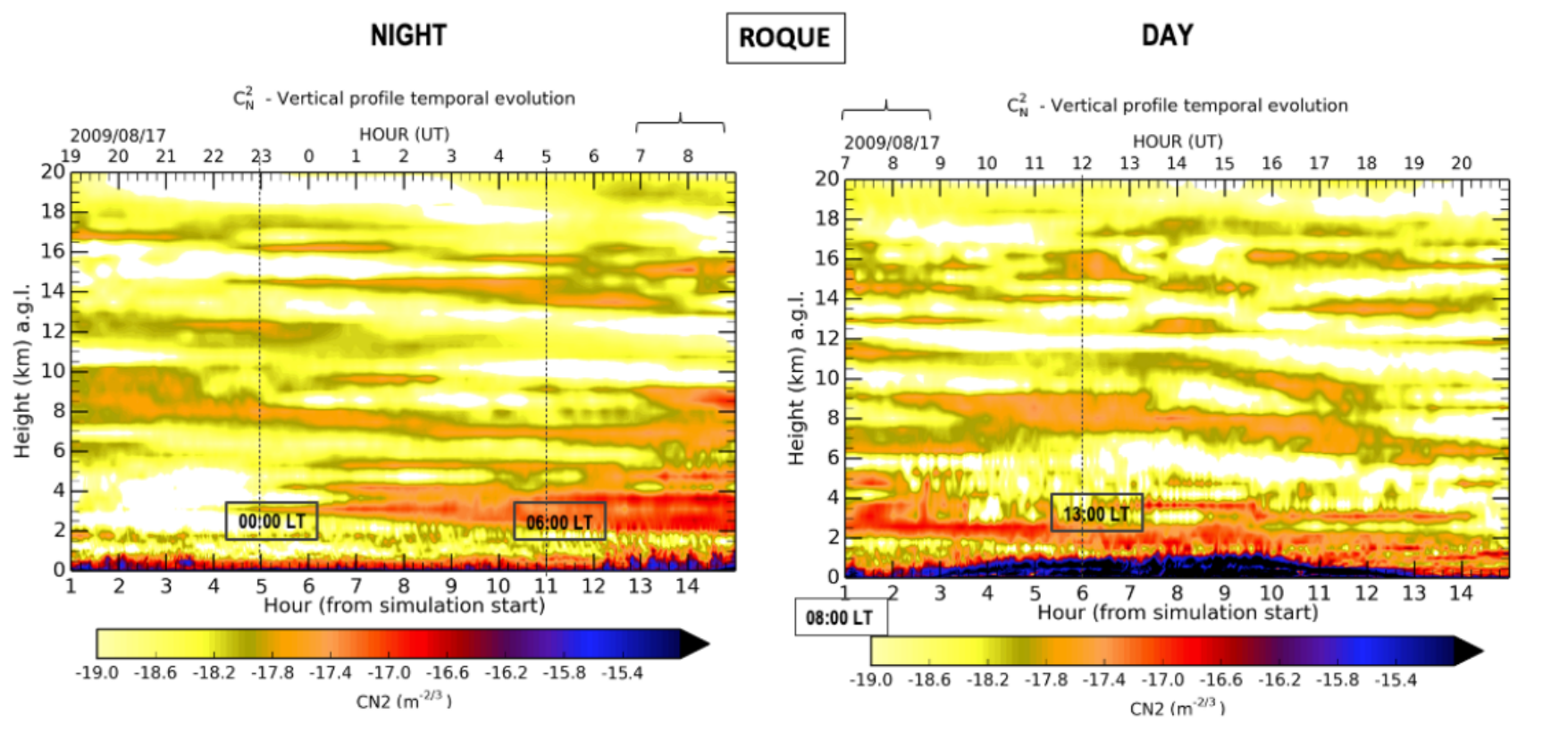}\\
\end{center}
\caption{\label{fig:fig4} Top: Temporal evolution of the $C_N^2$ profiles above the Roque de los Muchachos Observatory (ORM) related to the 2009/08/17 date during the night time (left) and during the day time (right). }
\end{figure*}

\begin{figure*}
\begin{center}
\includegraphics[angle=-90,width=0.8\textwidth]{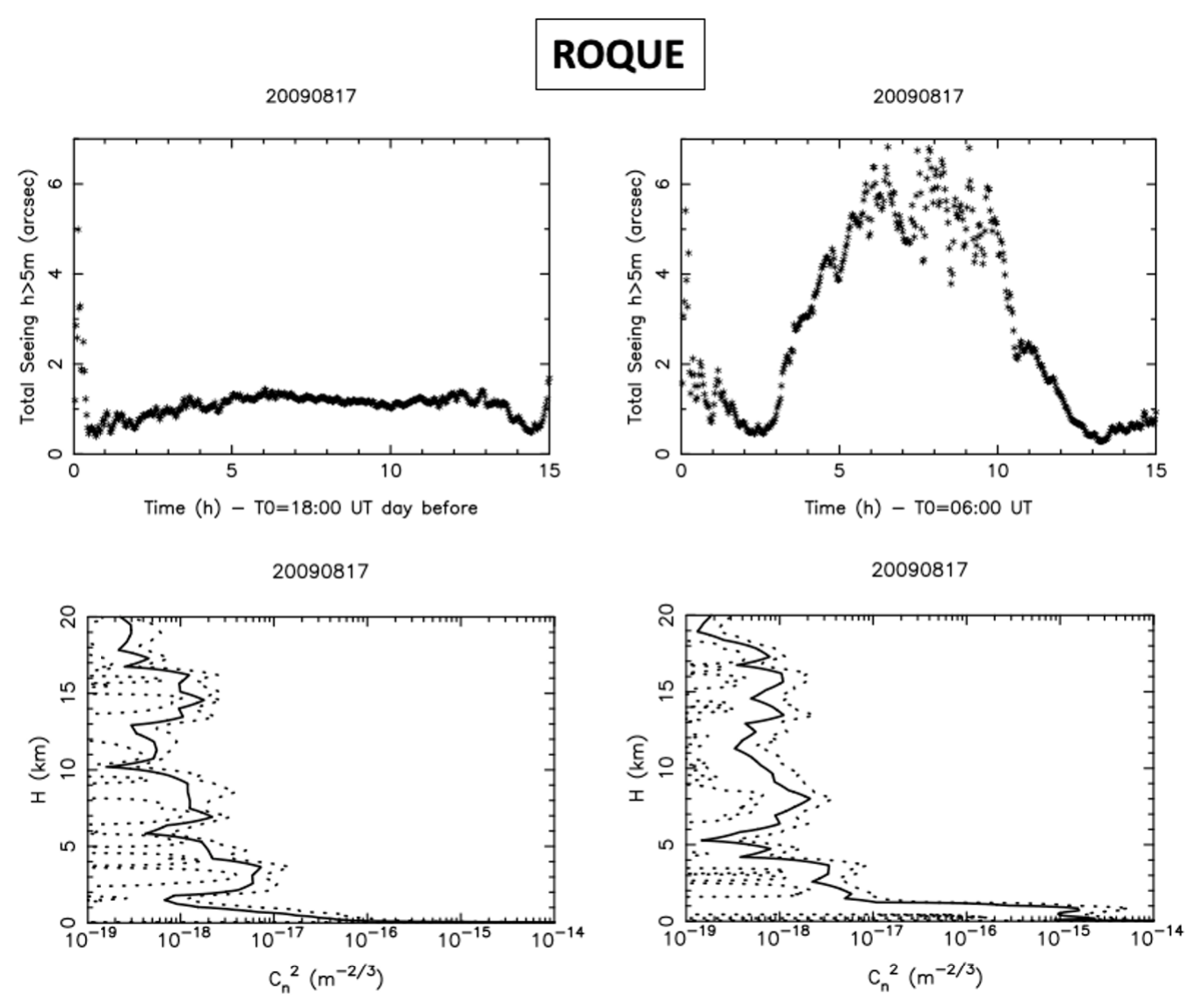}\\
\end{center}
\caption{\label{fig:fig5} Top: Temporal evolution of the seeing related to the date 2009/08/17 above the Roque de los Muchachos Observatory (ORM) during the night time (left) and during the day time (right). Bottom: average of the $C_N^2$ during the night (left) and during the day (right) of the same date. Dashed lines represent the standard deviation.}
\end{figure*}

\begin{figure*}
\begin{center}
\includegraphics[angle=-90,width=\textwidth]{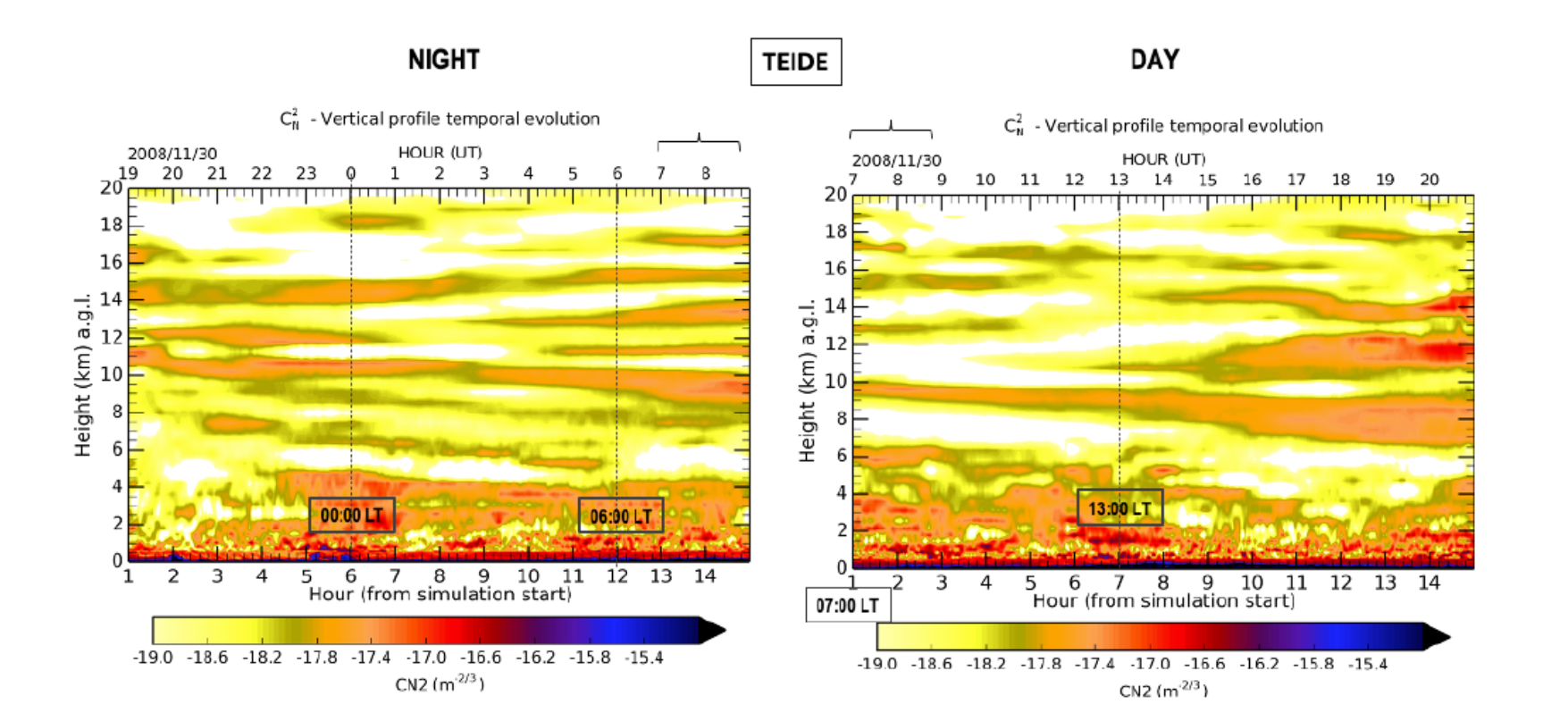}\\
\end{center}
\caption{\label{fig:fig6}  Top: Temporal evolution of the $C_N^2$ profiles above the Teide Observatory (TO) related to the 2008/11/30 date during the night time (left) and during the day time (right).}
\end{figure*}

\begin{figure*}
\begin{center}
\includegraphics[angle=-90,width=0.8\textwidth]{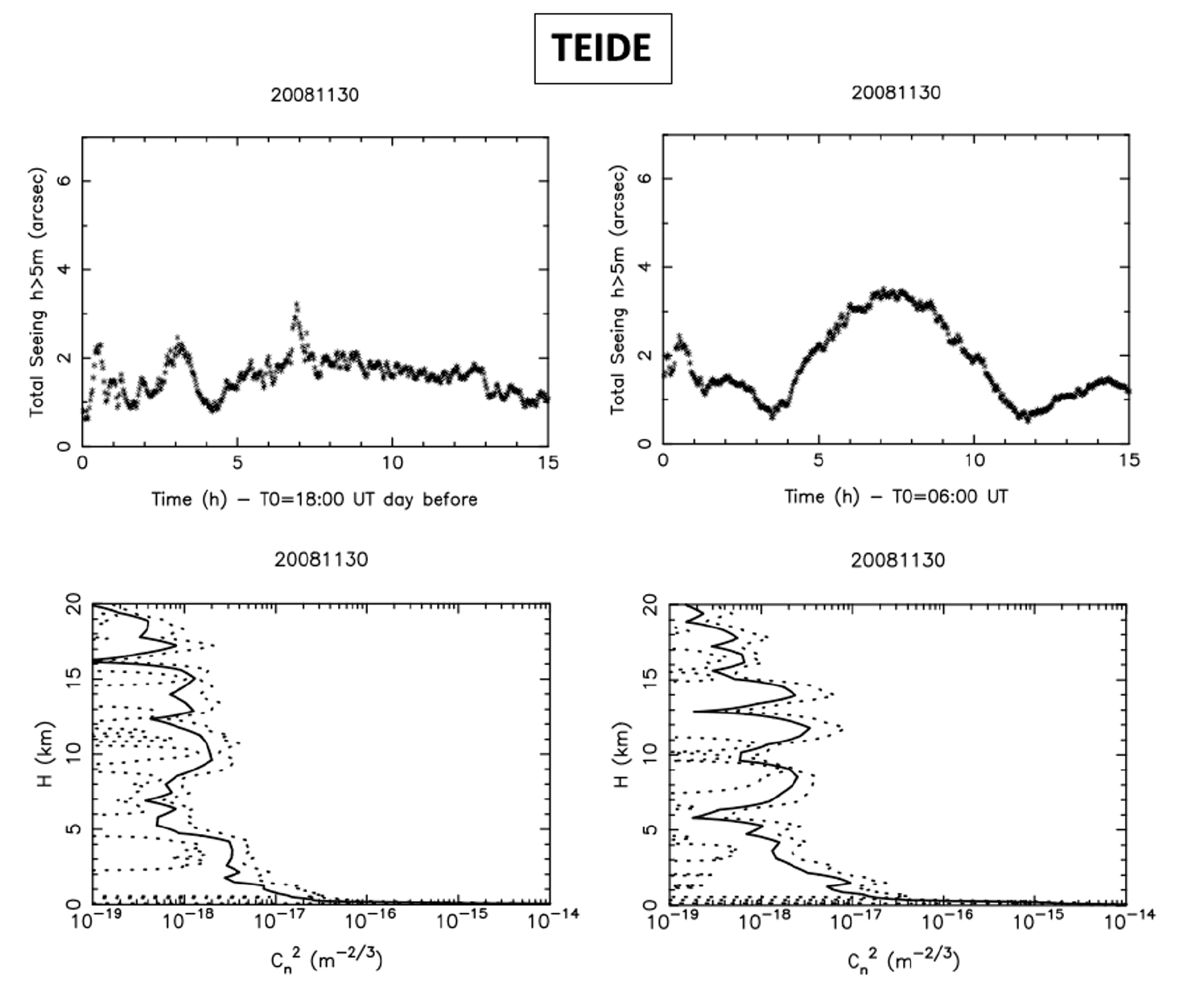}\\
\end{center}
\caption{\label{fig:fig7}  Top: Temporal evolution of the seeing related to the date 2008/11/30 above the Teide Observatory (TO) during the night time (left) and during the day time (right). Bottom: average of the $C_N^2$ during the night (left) and during the day (right) of the same date. Dashed lines represent the standard deviation.}
\end{figure*}

\begin{figure*}
\begin{center}
\includegraphics[angle=-90,width=0.8\textwidth]{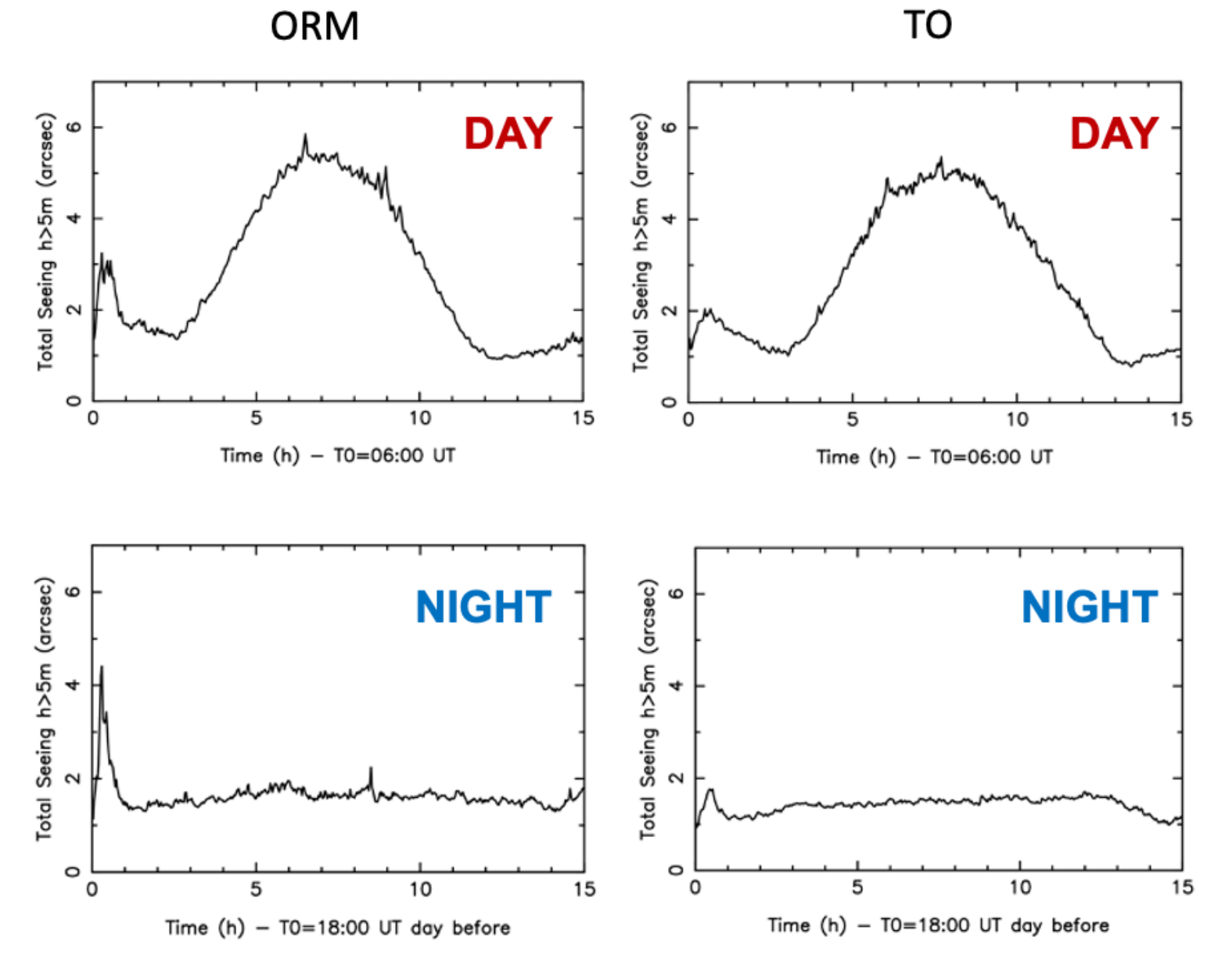}\\
\end{center}
\caption{\label{fig:fig8}  Top: Temporal evolution of the seeing during the day time averaged on the same 20 dates above ORM (left) and TO (right). Bottom: Temporal evolution of the seeing during the night time averaged on the same 20 dates shown previously above ORM (left) and TO (right).}
\end{figure*}

\section{PRELIMINARY RESULTS}

Here we compare the simulations performed with the atmospheric model during one day and the previous contiguous night and we estimate how the model reacts to the different conditions typical of the night and day time. We also perform a statistical analysis of results above both sites in night and day times, by using data from a sample of the same 20 dates for which we have observations above the two sites during the nights and days.

We refer to the scheme shown in Fig.\ref{fig:fig3} for the model configuration and the procedure used to perform a simulation. We refer the readers to \cite{masciadri2017,masciadri2020,masciadri2013} for details on the technical specifications of the mesoscale model Astro-Meso-Nh model and on the physical packages included. Figure \ref{fig:fig4} shows the $C_{N}^{2}$ temporal evolution during one night (2009/08/17) and the successive day above ORM. The convention used is that the date refers to the end of the simulation. For this reason, the date is the same for the night time and the day time. On the x-axis we report the UT time (top) and the time elapsed from the starting time of the simulation (bottom). 

As indicated in Fig.\ref{fig:fig3} the simulation related to the night covers the period [18:00 – 09:00] UT and the simulation related to the day covers the period [06:00 – 21:00] UT. The choice of the time in which to initialise the model depends on many factors including the time zone. First, we have to consider that initialisation data are calculated by the ECMWF model at synoptical hours (00:00, 06:00, 12:00, 18:00) UT. The beginning of a simulation has to be necessarily at one of these hours. At the same time, the beginning of the simulation has to be close to the beginning of the period we want to simulate (in our case the length of the ‘day’ and of the ‘night’) to optimise the simulation. 
Depending on the time zone of the astronomical sites, the interval between the beginning of the simulation and the beginning of the period that we want to reconstruct is of the order of a few hours. Considering the time zone of Canary Islands, as indicated in Fig.\ref{fig:fig3}, if we are interested in simulating the day time there is typically 1 or 2 hours (depending on the season) between the beginning of the simulation and the beginning of the period we are interested on. These characteristics are typical for this selected forecast scheme.

On the other side we have to remember that, when the simulation starts, the model needs some time to adapt itself to the ground and to the atmospheric initial conditions. This time, which is a sort of spurious time, is called spin-up time. Simulation outputs related to the spin-up time are usually rejected. It has been verified that the spin-up the model in the case of the Canary Islands is about 1 hour \cite{martelloni2018}. Fig.\ref{fig:fig4} shows the temporal evolution of the $C_{N}^{2}$ starting from 1 hour after the beginning of the simulation up to the end of the simulation.
Looking at Fig.\ref{fig:fig4}, the vertical dashed black lines report a few important time references expressed in local time (LT): on Fig.\ref{fig:fig4} (left panel) the central part of the night (00:00 LT) and the end of the night (at around 06:00 LT). On Fig.\ref{fig:fig4} (right panel) we have the central part of the day (13:00 LT) that corresponds to the peak of the radiation. Of course, the local time has to take into account the day light saving time. 

As the simulations for the night time and the day time are different we have a small interval of time of overlapped outputs (indicated by the graph parenthesis at top of each figure). Even if the continuity in the $C_{N}^{2}$ temporal evolution at the different heights is clearly visible, on the overlapped part we cannot have exactly the same identical vertical distribution because these outputs refer to different instants of time for the two simulations. In particular, on the figure on the left side we are at the end of the simulation of the night time; on the figure of the right side we are at the beginning of the simulation of the day time. 

On Fig.\ref{fig:fig4}-right, close to the ground, a bump of turbulence is well visible, clearly due to the solar radiation of the day time. On Fig.\ref{fig:fig5} we report the temporal evolution of the seeing during the night time (top-left) and during the day time (top-right) of the same sequence of night and day (2009/08/17) above ORM. The bump of seeing in the central part of the day is also well visible and coherent with what observed in Fig.\ref{fig:fig4}-right. In Fig.\ref{fig:fig5} we also show the average $C_{N}^{2}$ profile during the night (bottom-left) and during the day (bottom-right). Again, an excess of turbulence close to the ground is well visible in the $C_{N}^{2}$ profile (bottom-right).

In Fig.\ref{fig:fig6} and Fig.\ref{fig:fig7} we show results obtained for a night and day above the TO in a different date (2008/11/30). The same peak of turbulence close to the ground in correspondence of the day time is visible even if much weaker than in Fig.\ref{fig:fig4} and Fig.\ref{fig:fig5}. This is not relevant as the date is different but it put s in evidence the same reaction of the model during the day time. It is worth recalling that these results have been obtained from an uncalibrated model therefore it might make no sense to retrieve any quantitative conclusion from them on the level of turbulence distributed above ORM and TO. Nevertheless, they show that the model is able to reconstruct, for both sites, the expected distribution of turbulence from a qualitative point of view. We observe indeed that, in both cases, the diurnal trend is characterised by different features close to the ground with respect to the night. A more intense production of turbulence in coincidence with the local noon when the solar radiation is at its maximum is visible. In presence of solar radiation the turbulence is in a convective regime and the turbulence development is maximum in coincidence with the maximum solar radiation i.e. noon.  We also note that, in both cases, the turbulence in the free atmosphere evolves consistently between the night time and the day time. We observe that the local times of Fig.\ref{fig:fig4} and Fig.\ref{fig:fig6} are different because of the daylight saving time.

Even if the atmospheric model employed for the simulations presented above is not calibrated, we can use the rich sample of measurements available above both sites to perform a preliminary statistical analysis of results from atmospheric model simulations above ORM and TO. At this stage our goal is not that of comparing forecasts with measurements as the model is not calibrated but we can, however, perform an analysis in relative terms of the integrated values between the two sites. 
Fig.\ref{fig:fig8} shows the temporal evolution of the seeing calculated over 20 dates in the night and day time above ORM and TO. We calculated here the average of the J values related to the 20 dates and then we retrieved the seeing following the relation:

\begin{equation}
\varepsilon=1.99754 \cdot 10^7 \cdot J^{\frac{3}{5}}
\end{equation}

The 20 dates belong to the site testing campaigns of GLF\cite{garcia-lorenzo2011} and GLFb\cite{garcia-lorenzo2011b}. 
Fig.\ref{fig:fig8} shows that there are no major differences on the seeing developed above ORM and TO during the day time that followed the nights of the analysed data. Also, the turbulence developed during day time is almost 3 time larger than that modelled during the night time. Having said that, it is important to remember that the model is not calibrated therefore the analysis can be done only in relative terms. A more accurate estimate should be done with a calibrated model. Indeed the difference between night and day might be modified in quantitative terms with a calibrated model but it remains valid the conclusion that during the day time the turbulence is more consistent than during the night time. The interesting result is that, the comparative analysis is poorly dependent from the calibration therefore the fact that ORM and TO are comparable is a convincing confirmation that ORM can be a good choice for the EST (and, at least, it was not a bad choice).

Initialisation and forcing data come from the HRES model is the General Circulation Model of the European Centre for Medium-Range Weather Forecasts (ECMWF) used for the daily real time forecasts characterised by a horizontal resolution of 9-10 km. This is the highest resolution for this kind of models. 
We prefer to use initialisation data coming from HRES model (and not the ERA) data as we are interested in treating the problem of automatic forecasts in operative configuration.

\section{OPTICAL TURBULENCE MEASUREMENTS DURING DAYTIME}

As already mentioned the Astro-Meso-Nh model in this study is not calibrated. We summarise here briefly the concept of the model calibration for the optical turbulence. The reconstruction of the $C_{N}^{2}$ done with a mesoscale model dedicated to the optical turbulence such as the Astro-Meso-Nh model requires a model calibration for a correct quantitative estimate. The calibration is a procedure that aims to fix a free parameter in the model and that can provide reliable quantitative estimates. Briefly, the model depends on a free parameter called turbulent kinetic energy (E$_{min}$) that has to be fixed through a fitting with observations. The first method aver proposed for a model calibration applied to the $C_{N}^{2}$ was MJ2001\cite{masciadri2001} which proposed a strategy to fix the E$_{min}$ by using a rich statistical sample of $C_{N}^{2}$ vertical profiles observed above the site of interest. The method is based on the fact that it is indeed possible to prove that, under condition of thermodynamic stability, the $C_{N}^{2}$ is proportional to the E$_{min}$. Starting from this method, further methods have been implemented in atmospheric models later on such as H2011\cite{hagelin2011} and M2017\cite{masciadri2017}. In some cases, the use of integrated measurements is used together with vertical $C_{N}^{2}$ profiles to perform the model calibration\cite{masciadri2017}. The important thing to retain is that, to perform a model calibration, vertical stratification of the $C_{N}^{2}$ extended on the whole 20 km related to a statistical sample that is as rich as possible is necessary.

\begin{figure*}
\begin{center}
\includegraphics[width=0.4\textwidth]{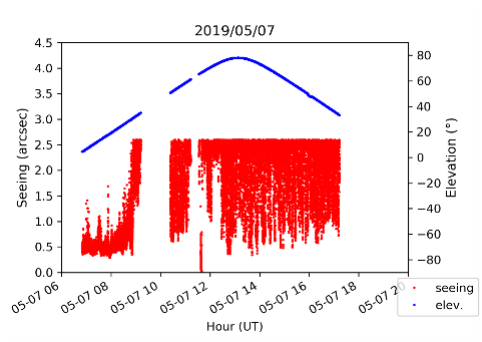}\\
\end{center}
\caption{\label{fig:fig9}  Temporal evolution of the seeing measurements obtained with the WF-WFS during one night. Red line is the raw seeing, the blu line represents the line of across the sky of the telescope.  }
\end{figure*}

Observations of the optical turbulence that we collected and/or took into consideraiton in the context of the SOLARNET project during the day time revealed to be not really suitable for a study aiming to calibrate the Astro-Meso-Nh model for our goal. They are:\\
\begin{itemize}
\item SHABAR measurements \\
\noindent
We analysed measurements collected for long-time by IAC above ORM and TO in the past\cite{solarnet_1,solarnet_2} but these measurements provide the turbulence stratification on the very shallow extension of a few hundred of meters above the ground. This is actually the part of the atmosphere in which the calibration is less effective. \\
\item WF-WFS measurements\\
\noindent
These are measurements are obtained thanks to a SDIMM+ technique\cite{scharmer2010} that is in principle able to provide a stratification of the turbulence up to around 12~km and the seeing. Such an instrument was supposed to run above ORM (at the telescope SST) and TO (at the telescope VVT). Unfortunately, for a set of technical reasons, these measurements could not be carried out and we could access only integrated measurements obtained above ORM (Goran Shermer - private communication). Our analysis, however, arrived to the conclusion that these measurement could not be used as they saturated at around 2.5” (Fig.\ref{fig:fig9}) or, to better said, a seeing $>$ 2.5" is not reliable (Goran Scharmer - private communication). That means that part of the turbulence spectrum is not reconstructed by this method. A maximum seeing of 2.5" means a $r_{0}$ $>$ 3.9~cm but we know that in reality we should expect a $r_{0}$ also smaller than this value during the day time\cite{socas2005}. Measurements of the $C_{N}^{2}$ were planned to be available but due to technical reasons they have never been performed. 
\end{itemize}

The efforts invested on instrumentation for estimates of optical turbulence stratification on the 20~km a.g.l. during the day time is more recent than that characterising the night time. It is evident the necessity of reliable instruments (preferably monitors) that might be used contonously during the day time. In the last years a few attempts have been done in this direction, for example, the PML\cite{chabe2020,tengfei2020}, and, more recently, the 24HSHIMM\cite{griffith2023}. PML is based on the observation of the sun edge and presents however some not negligible limitations. The most relevant is that important relative errors have been estimated for $C_{N}^{2}$ when h $>$ 1~km. The relative error increases with height and reaches $\approx$ 50\% for h $>$ 10~km\cite{aristidi2020}. The instrument 24HSHIMM is based on the observation of stars during the day time in short-wave infrared (SWIR) band and it has not the drawbacks of PML in this respect. Said that the impression is that these instruments have not yet achieved a required strengthens. A critical point for measurements of the turbulence stratification during the day time is that it is not trivial to perform a validation of the technics because of lack of references. 

\section{CONCLUSIONS}

In this paper we presented the preliminary analysis performed in the context of a long-term study aiming to transfer to the day time the method of the optical turbulence forecast that our group developed for the night time. We presented the digital elevation models of the ORM and TO, the model configuration, the forecast scheme and preliminary results. We could prove that the $C_{N}^{2}$ extended on the 20~km reconstructed by the Astro-Meso-Nh model above the ground during the day time shows a turbulence production close to the ground much more consistent with respect to the night time as expected. This is evident above the ORM and TO. Besides, we performed a forecast during the night and day time in relation to 20 dates above ORM and TO. Our preliminary analysis (see Fig.\ref{fig:fig8}) shows that there are no major differences in seeing above ORM and TO during the day time and the seeing at noon during the day time is almost 3 times larger than during the night. It is worth to highlight, however, that we treated an uncalibrated model therefore it is possible that, from a quantitative point of view, with a calibrated model, the difference of seeing between night and day time at the point of maximum difference might be weaker or higher (i.e. the quantitative estimate can be change) but, in any case, this does not affect the fact that the seeing during the day time is larger than during the time time. The time of maximum difference corresponds to the time of maximum solar radiation therefore around 12:00-13:00 LT. In this paper we have shown that on a statistic of 20 dates, ORM and TO show comparable seeing estimate at the time of the maximum solar radiation. This is a good news for EST as the site of ORM has been in any case already established to be the site of EST.
The progresses in our studies depend strongly on the availability and reliability of observations during the day time. We think that this is the roadmap to follow.

\acknowledgments 
 
The study has been co-funded by the FCRF foundation through the 'Ricerca Scientifica e Tecnologica' action - N.45103. by the project ALTA Center (ENV001, ENV002) and by the European Union's Horizon 2020 research and innovation programme under grant agreement No 824135 (SOLARNET). Initialisation and forcing data of the Astro-Meso-Nh model come from the HRES general circulation model of the ECMWF. We acknowledge Bego\~na Garcia-Lorenzo, Julio Castro Almazan, Casiana Mu\~noz-Tu\~non, Luz Maria Montoya Martinez from IAC, Albar Garcia de Gurturbai Escudero from TNG, Neil O'Mahony from ING, Goran Sharmer from US for sharing observations of the optical turbulence related to the two astronomical sites ORM and TO.

\bibliographystyle{spiebib} 

\end{document}